\documentclass[12pt,a4paper ]{article}

\newcommand{\frat}[2]{\frac{\textstyle #1}{\textstyle #2}}
\newcommand{\vf}[1]{\mbox{\boldmath $#1$}}

\begin{document}

\begin{center}{\Large
\bf Towards possible origin of binding in non-abelian gauge
theories}
\\
\vspace{0.5cm} S.V. Molodtsov, G.M. Zinovjev$^\dagger$
\\
 \vspace{0.5cm}{\small\it State Research
 Center, Institute of Theoretical and Experimental Physics, 117259,Moscow, RUSSIA}
\\ $^\dagger$ {\small \it Bogolyubov Institute for
Theoretical Physics,
\\ National Academy of Sciences of Ukraine,UA-03143, Kiev, UKRAINE}
\end{center}
\vspace{0.5cm}
%\begin{abstract}
\begin{center}
\begin{tabular}{p{16cm}}
{\small{Estimate of average energy of the non-abelian dipole in
the instanton medium is calculated. This energy for the dipole in colour
singlet state escalates linearly with the separation
increasing for the point-like sources
unscreened. And its 'tension' coefficient develops the magnitude pretty similar
to one as inferred from the lattice QCD and other model approaches.}}
\end{tabular}
\end{center}
%\end{abstract}
\vspace{0.5cm}
PACS: 11.15 Kc, 12.38-Aw

Recent evolution scenario for the QCD matter being
created in ultrarelativistic heavy ion collisions proposes
interesting explanation of origin of longstanding 'mixed phase'
which was found in lattice calculations of QCD at finite
temperatures and, seems, is experimentally observed \cite{0}. A
key point to account for the phenomenon in Ref.\cite{0} is to
justify the presence of colour Coulomb binding interaction and
instanton induced interaction (hence, a lot of bound states
available) in rather wide temperature region above the critical
temperature. In spite of many intuitive inputs the ideas of this
scenario (together with some hints at colour singlet states
available everywhere in the coupling space in Ref.\cite{00}) may
help in gaining insight on the QCD dynamics. Our major concern
here is to answer question whether the origin of those bound
states approaching the critical temperature is instructive to
understand the nature of ordinary hadrons.

Since the discovery of
(anti-)instantons (the classical solutions to the Yang-Mills
equations with nontrivial topological features) the problem of the
gauge interaction between instantons is under intensive study and
the dipole interaction of those pseudo-particles with an external
field has already been argued in the pioneering papers \cite{1}.
This very important conclusion was grounded on examining the reaction
of a single instanton while in a weak slowly varying (constant)
external field at distances compared to the instanton scale size. We
follow that standard approach in considering the approximate solution
of the Yang-Mills equations as
\begin{equation}
\label{1}
{\cal A}^a_{\mu}(x)=A^a_{\mu}(x)+B^a_{\mu}(x)~.
\end{equation}
Here the first term denotes the field of a single (anti-)instanton
in the singular gauge as we are planning to consider an
(anti-)instanton ensemble and need to have the non-trivial
topology at the singularity point
\begin{equation}
\label{inst}
A^a_{\mu}(x)=\frat2g~\omega^{ab}\bar\eta_{b\mu\nu}~\frat{\rho^2}{y^2+\rho^2}~\frat{y_\nu}{y^2}~,
\end{equation}
where $\rho$ is an arbitrary parameter characterizing the
instanton size centered at the coordinate $z$ and colour
orientation defined by the matrix $\omega$, $y=x-z$,
$\bar\eta_{b\mu\nu}$ is the 't Hooft symbol (for anti-instanton
$\bar\eta \to \eta$) and $g$ is the coupling constant of non-abelian
field. The second term describes an external field. In fact, the result of
Ref. \cite {1} as to the instanton interaction with external field (the
CDG term) may be obtained in more general way by calculating
the corresponding effective Lagrangian (without specifying the external
field characteristics) \cite {sergey}. For the sake of simplicity we
limit ourselves here to dealing with $SU(2)$ group only and begin
studying an external field originated by immovable Euclidean colour
point-like source $e\delta^{3a}$ and Euclidean colour dipole $\pm
e\delta^{3a}$ which is one of two configurations only providing with the
static solutions. Fixing the location of point-like sources is gauge
invariant unlike specifying their orientation in 'isotopic' space.
In order to get rid of this 'vice' \cite{sergey1} the solution
should be characterized by its energy and total 'isospin'
('isospin' of source plus 'isospin' carried by the Yang-Mills
field) because these quantities are gauge invariant in addition to
being conserved.

With this 'set-up' we could now claim that the
results obtained before have maintained the interacting terms
proportional $e/g$ only and in this paper we aim to analyze the
contributions proportional $e^2$ which display rather indicative
behaviour of asymptotic energy of Euclidean non-abelian point-like
sources while in the instanton liquid (IL) \cite{2} and have nothing
to do with the additional corrections due to the quantum fluctuations
about the saddle point. We treat the potentials in their Euclidean forms
and then the following changes of the field and Euclidean point-like
source variables are valid $B_0\to i B_4$, $e\to -i e$ at transition
from the Minkowski space. Actually, last variable change is resulted
from the corresponding transformations of spinor fields
$\psi\to\hat\psi$, $\bar\psi\to-i\hat\psi^\dagger$,
$\gamma_0\to\gamma_4$. It means we are in full accordance with
electrodynamics where the practical way to have a pithy theory in
Euclidean space is to make a transition to an imaginary charge.
Surely, these Euclidean sources generate the
fields of the same nature as ones we face taking into account gluon
field quantum fluctuations $a_{qu}$ around the classical instanton
solutions $A=A_{inst}+a_{qu}$.

For Euclidean point-like source and
dipole in colourless state we take the Coulomb solutions of the
classical Yang-Mills equations in the following forms
\begin{eqnarray}
\label{coulomb} && B^a_{\mu}(x)=({\vf 0},\delta^{a3}~\varphi),~
\varphi=\frat{e}{4\pi}~\frat{1}{|{\vf x}-{\vf z}_e|}~,\nonumber\\
[-.2cm]\\ [-.25cm] &&B^a_{\mu}(x)=({\vf
0},\delta^{a3}~\varphi),~\varphi=\frat{e}{4\pi}~\left(\frat{1}
{|{\vf x}-{\vf z}_1|}-\frat{1}{|{\vf x}-{\vf z}_2|}\right)~,\nonumber
\end{eqnarray}
respectively, with ${\vf z}_e$ being a coordinate of point-like
source in peace and ${\vf z}_1$, ${\vf z}_2$ as coordinates
characterizing a dipole. As known, the field originated by
particle source in 4d-space develops the cone-like shape with the
edge being situated just at the particle creation. Then getting
away from that point the field becomes fully developed and steady in
the area neighboring to it. In distant area where the field penetrates
into the vacuum it should be described by the retarded solution obeying,
in particular, the Lorentz gauge $\partial_\mu B^a_{\mu}=0$ (which is
also valid for Eq.(\ref{inst})). Here we are interested in
studying the interactions in the neighboring field area and the
cylinder-symmetrical Coulomb field (which is $x_4$ independent as
in Eq.(\ref{coulomb})) might be its relevant image in 4d-space.

Thus, the field strength tensor is given as
\begin{equation}
\label{2}
G_{\mu\nu}^a=\partial_\mu {\cal A}^a_{\nu}-\partial_\nu{\cal A}^a_{\mu}+
g~\varepsilon^{abc}{\cal A}^b_{\mu}{\cal A}^c_{\nu}~,
\end{equation}
where $\varepsilon^{abc}$ is the entirely asymmetric tensor and for the
field superposition of Eq.(\ref{1}) it can be given by
\begin{eqnarray}
\label{3}
&&G_{\mu\nu}^a=G_{\mu\nu}^a(B)+G_{\mu\nu}^a(A)+G_{\mu\nu}^a(A,B)~,\nonumber\\
[-.2cm]\\
[-.25cm]
&&G_{\mu\nu}^a(A,B)=g~\varepsilon^{abc}(B^b_{\mu}A^{c}_\nu+A^b_{\mu}B^{c}_\nu )~,\nonumber
\end{eqnarray}
if $G_{\mu\nu}^a(A)$ and $G_{\mu\nu}^a(B)$ are defined as in
Eq.(\ref{2}). Then gluon field strength tensor squared reads
\begin{eqnarray}
\label{4}
 G_{\mu\nu}^a~ G_{\mu\nu}^a
 &=&G_{\mu\nu}^a(B)~G_{\mu\nu}^a(B) + G_{\mu\nu}^a(A)~G_{\mu\nu}^a(A) +
G_{\mu\nu}^a(A,B)~G_{\mu\nu}^a(A,B) +\nonumber\\
 [-.2cm]
\\[-.25cm]
&+&2~ G_{\mu\nu}^a(B)~G_{\mu\nu}^a(A) + 2~G_{\mu\nu}^a(B)~G_{\mu\nu}^a(A,B) +
 2~G_{\mu\nu}^a(A)~G_{\mu\nu}^a(A,B)~. \nonumber
\end{eqnarray}
The various terms of Eq.(\ref{4}) provide the contributions of
different kinds to the total action of full initial system of
point-like sources and pseudo-particle
\begin{equation}
\label{5}
S=\int dx \left( \frat14~ G_{\mu\nu}^a~G_{\mu\nu}^a+j^a_{\mu}{\cal A}^a_{\mu}\right)=
S_e(B)+\beta+S_{int}~.
\end{equation}
So, the first and second terms of Eq.(\ref{4}) account for the source
self-energy $S_e$ (for the dipole it should be supplemented by the
Coulomb potential of interacting sources) and the single instanton
action $\beta=8\pi^2/g^2$. At small distances the first term is
regularized by introducing the source 'size' as in classical electrodynamics.
These terms are proportional to $e^2$ and $g^{-2}$, respectively and,
clearly, out of our interest here. The fourth and last terms of Eq.(\ref{4})
together with the term $j^a_{\mu}A^a_{\mu}$ of Eq.(\ref{5}) account for the
contribution to the interacting part of the action $S_{int}$ proportional
to $e/g$. And the third (repulsive) term of Eq.(\ref{4})
only leads to the contribution proportional to $e^2$ as the fifth term is
equal zero owing to the gauge choice. Collecting those non-zero
contributions we have after performing the calculations the following
result
\begin{equation}
\label{6}
S_{int}=\frat{e}{g}~\bar\eta_{k4i}~\omega^{3k} I_i+\left(\frat{e}{4\pi}\right)^2
J+ \left(\frat{e}{4\pi}\right)^2 K_{kl}~\omega^{3k}\omega^{3l}~.
\end{equation}
The exact form of $I_i$ (where the CDG term is absorbed) is unnecessary in
this paper because going to relate an external field to the stochastic field
of fluctuations (the IL model) we have to average over the colour
orientation of (anti-)instanton. It leads to disappearance of the dipole
contribution and the first nonvanishing correction, as known, comes from a
condensate term (but in next order). Two other terms result from the
'mixed' component of field strength tensor which looks like
\begin{eqnarray}
\label{mc}
&G_{4i}^a(A,B)=2~\frat{e}{4\pi}~\varepsilon^{a3c}~\omega^{ck}~\bar\eta_{ki\alpha}
~\frat{y_\alpha}{y^2}~\frat{\rho^2}{(y^2+\rho^2)}~\frat{1} {|{\vf
y}+{\vf \Delta}|}~, &\nonumber\\ [-.2cm]
\\[-.25cm]
&G_{4i}^a(A,B)=2~\frat{e}{4\pi}~\varepsilon^{a3c}~\omega^{ck}~\bar\eta_{ki\alpha}~
\frat{y_\alpha}{y^2}~\frat{\rho^2}{(y^2+\rho^2)}~\left(\frat{1}{|{\vf y}+{\vf \Delta}_1|}
-\frat{1}{|{\vf y}+{\vf\Delta}_2|}\right)~,&\nonumber
\end{eqnarray}
for the point-like source and for the field of colourless dipole,
respectively, and where ${\vf \Delta}={\vf z}-{\vf z}_e$, ${\vf
\Delta}_{1,2}={\vf z}- {\vf z}_{1,2}$. The other contributions to
$G^a_{\mu\nu}(A,B)$ are absent because of the gauge used. In order
to handle the further formulae easily it is practical to introduce
new dimensionless coordinates as $x/\rho\to x$. Then for the
single source the function $J$ and the tensor $K$ take the
following forms $$J=2~\int dy~\frat{2~y^2-{\vf
y}^2}{y^4~(y^2+1)^2~|{\vf y} + {\vf \Delta}|^2}~,$$
$$K_{kl}=2~\int dy~ \frat{y_ky_l}{y^4}~\frat{1}{(y^2+1)^2~|{\vf
y}+ {\vf \Delta}|^2}~,$$ and can not be integrated in the
elementary functions. Fortunately, we need  to know their asymptotic
values at $\Delta\to\infty$ only
$$J\simeq\frat{5\pi^2}{2}\frat{1}{\Delta^2}~,$$ and for the
components of the 2-nd rank tensor
$$K_{ij}=\delta_{ij}~K_1+\hat\Delta_i\hat\Delta_j~K_2~,$$ we have
$$K_1\simeq\frat{\pi^2}{2}\frat{1}{\Delta^2}~,~~K_2\simeq 0~.$$
Clearly, the 'mixed' component of field strength tensor is of purely
non-abelian origin but its contribution to the action of whole
system (point-like sources and pseudo-particle) takes the form of
self-interacting Euclidean source $\sim e^2$ although it is
descended from the instanton field and field generated by sources.
It seems, this simple but still amazing fact was not explored
properly.

Now we are trying to analyze the pseudo-particle
behaviour in the field of Euclidean non-abelian source and develop
the perturbative description related to the pseudo-particle
itself. Realizing such a program we 'compel' the pseudo-particle
parameters to be the functions of 'outside influence', i.e.
putting $\rho\to R(x,z)$, $\omega\to\Omega(x,z)$. These new
fields-parameters are calculated within the multipole expansion
and then, for example, for the (anti-)instanton size we have
\begin{eqnarray}
\label{dec}
R_{in}(x,z)&=&\rho+c_\mu~y_\mu+c_{\mu\nu}~y_\mu~y_\nu+
\dots~,~~~~~|y|\leq D \nonumber\\
[-.2cm]
\\[-.25cm]
R_{out}(x,z)&=&\rho+d_\mu~\frat{y_\mu}
{y^2}+d_{\mu\nu}~\frat{y_\mu}{y^2}~\frat{y_\nu}{y^2}+\dots~,
~~~|y|>D~, \nonumber
\end{eqnarray}
Similar expressions could be written down for the (anti-)instanton
orientation in the colour space $\Omega(x,z)$ with $D$ being a certain
parameter fixing the radius of sphere where the multipole expansion
growing with the distance increasing should be changed for the decreasing
one being a result of deformation regularity constraint imposed. Then the
coefficients of multipole expansion $c_\mu,c_{\mu\nu},\dots$ and
$d_\mu,d_{\mu\nu},\dots$ are the functions of external impact. It turns
out this approach allows us to trace the evolution of approximate solution
for the deformed (crumpled) (anti-)instanton as a function of distance
and, moreover, to suggest the self-consistent description of pseudo-particle
and point-like source fields within the approximate solution of Eq.(\ref{1})
(our paper has been completed recently). The role of deformation fields
was found to be essential in the local vicinity of source (at distances of
order $\leq 2\rho$) and results in considerable growth of the pseudoparticle
action.

Proceeding to the calculation of average energy of point-like source
immersed into IL we have to remember that one should work with the
characteristic configuration which is the superposition of instanton and
anti-instanton fields also supplemented by the source field
$B^a_{\mu}(x)$, i.e.
\begin{equation}
\label{8}
 {\cal A}^{a}_\mu(x)=B^a_{\mu}(x)+\sum_{i=1}^NA^{a}_\mu(x;\gamma_i)~,
\end{equation}
where  $\gamma_i=(\rho_i,z_i,\omega_i)$ are the parameters
describing the $i$-th (anti-)instanton.  This would-be superposition
ansatz is a solution to the Yang-Mills equations only in the limit of
infinite separation but approximates well enough the dominant
configuration saturating the functional integral \cite{1}. The
IL density at large distances from source practically coincides with its
asymptotic value $n(\Delta)\sim n_0 e^{\beta-S}\simeq n_0$ because the
action of any pseudo-particle here is approximately equal $\beta$. The
quantity we are interested in should be defined by averaging $S$
over the pseudo-particle positions and their colour orientations
(taking all pseudo-particles of the same size at the moment) as
\begin{eqnarray}
\label{s}
 &&\langle S\rangle=\prod_{i=1}^N \int\frat{d z_i}{V}~\int d\omega_i~~S=\nonumber\\
 [-.2cm]
\\[-.25cm]
&&=\frat{e^2}{4\pi}\frat{1}{a}~X_4+N~\beta+N\int \frat{d {\vf \Delta}}
{L^3}\left(\frat{e}{4\pi}\right)^2
\left(J+\frat{K_{ii}}{3}\right)~,\nonumber
\end{eqnarray}
here $L$ is a formal upper integration limit, $V=L^3X_4$ defines the IL
volume, $X_4$ is an upper bound of the 'time' integration, $N$ denotes the
total number of pseudo-particles and $a$ is a source 'size' value (on
strong interaction scale, of course). Taking the asymptotic functions of
$J$ and $K$ and returning to the dimensional variables for the moment one
instanton contribution to the average action can be written down in the
following form
\begin{equation}
\label{sav} \langle S\rangle\simeq\frat{e^2}{4\pi}\frat{1}{a}~X_4+
\frat{N}{V}~\beta~L^3 X_4+\frat{N}{V}\frat{6 \pi^3}{\beta}
\frat{e^2}{g^2}~\bar\rho^2~L~X_4~,
\end{equation}
where $\bar\rho$ is the mean size of (anti-)instanton in IL. The result is
given in the form with the common factor $X_4$ extracted because our concern
now is the (anti-)instanton behaviour in the source field background where
the field is fully developed and the solution possesses the scaling
property at any time slice. In the limit $N,V\to\infty$ and at the IL
density $n=N/V$ fixed this result becomes
\begin{equation}
\label{14}
E\simeq \frat{e^2}{4\pi}\frat{1}{a}+
n~\beta~L^3 +n~\frat{6 \pi^3}{\beta} \frat{e^2}{g^2}~\bar\rho^2~L~.
\end{equation}
The last term here looks like a correction to the gluon condensate (the
second term). However, this contribution linearly
increasing with $L$ and proportional to $e^2$ has different physical
meaning of additional contribution to the source self-energy
\begin{equation}
\label{dd}
E\simeq \sigma~L~,~~~\sigma=n~\frat{6 \pi^3}{\beta}\frat{e^2}{g^2}~\bar\rho^2~.
\end{equation}
From the view point of the CDG 'ideology' \cite{1} an appearance of such a
term is quite understandable if we remember that for long-range Coulomb
solutions Eq.(\ref{coulomb}) the tensor $G_{\mu\nu}^a$ falls like
$r^{-2}$ at large distances (similar to merons and regular instantons), 
markedly slower than one for the instanton
($\sim r^{-4}$). Besides, the colour 'magnetic field' behaviour for our
configuration (of time slice) with point-like source looks as
$$H_{k}^a=-\delta^{a3}~\frat{x_k}{|{\vf x}|^3}~,$$
bringing the result to the formal analogy with the Polyakov's model of
vacuum at a fixed time albeit its physical meaning here is much more
obscure and rather far from the interpretation discussed in \cite{1}.

The terms next to leading order could be obtained by a cluster
decomposition of the exponential with the interacting term derived, i.e.
Eq.(\ref{6})
\begin{equation}
\label{dop} 
\langle \exp (-S_{int})\rangle_{\omega z}=\exp
\left(~\sum_k\frat{(-1)^k}{k!}~\langle\langle S^k_{int}\rangle
\rangle_{\omega z}\right)~,
\end{equation}
where the indecies $\omega z$ just imply aforementioned averaging over the
pseudo-particle positions and their colour orientations and
$\langle S_1\rangle=\langle\langle S_1\rangle\rangle$, 
$\langle S_1 S_2\rangle=\langle S_1\rangle\langle S_2\rangle+
\langle\langle S_1 S_2\rangle\rangle, \dots$
Then, for example, the second order contribution generated by dipole interaction 
$I_i$ looks like
$$\langle\langle \left( S^{(I)}_{int}\right)^2\rangle\rangle=~
\frat{e^2}{g^2}~\frat{8\pi^3}{3}~\left(\pi-\frat{8}{3}\right)~\frat{\bar\rho^3}{L^3}~,$$
and results in (due to nontrivial dielectic IL properties \cite{1}) the regular
correction to the average IL energy (\ref{14}) as
\begin{equation} 
 \Delta E^{(I)} = -\frat{N}{V}\frat{e^2}{g^2}~\frat{4\pi^3}{3}~
\left(\pi-\frat{8}{3}\right)~\bar\rho^3 ~.
\end{equation}

Now dealing with the field of colour dipole in the colour singlet
state we are able to demonstrate the resembling reaction of IL
once more. The contribution to average energy $E$ is defined at
large distances by the integral very similar to that for the
configuration with one single source
\begin{equation}
I_d=\int d{\vf \Delta}_1 \left(\frat{1}{{\vf\Delta}_1^{2}}
 -2~\frat{1}{|{\vf\Delta}_1| |{\vf\Delta}_2|}+\frat{1}{{\vf\Delta}_2^{2}}\right)~.
\end{equation}
When the source separation $l=|{\vf z}_1-{\vf z}_2|$ is going to
zero, the field disappears and the final result should be zero.
There are only two parameters  to operate with the integral, they
are $L$ and $l$. The dimensional analysis reveals that  the
integral is the linear function of both but the $l$-dependence
only obeys the requirement of integral petering out at $l\to 0$.
It is easy to receive the equation
$$\frat{I_d}{4\pi}=L-2\left(L-\frat{l}{2}\right)+L=l~$$ for
determining the coefficient (the contributions of three integrals
are shown separately here). Finally the average dipole energy
for the colour singlet (c.s.) state reads
\begin{equation}
E\simeq \sigma~l~,~~~{\mbox (c.s.)}~,
\end{equation}
The 'tension' value $\sigma$ for
the IL characteristic parameters $\frat{\bar\rho}{\bar R}\simeq
\frat{1}{3}$ where $\bar R$ is the mean separation of
pseudo-particles, $n=\bar R^{-4}$, $\beta\simeq 12$,
$\bar\rho\simeq 1$ GeV$^{-1}$ \cite{2},\cite{3} comes about
$\sigma \simeq0.6$ GeV/fm (if one takes for the brief estimate for
the source intensity $e\simeq g$). This result looks fully
relevant in view of the qualitative character of estimates which
IL is able to provide with. Moreover, such a value is in
reasonable agreement with the estimates extracted from the
potential models for heavy quarkonia, for example. If one intends
to explore the magnitudes like $\langle
S_{int}~e^{-S_{int}}\rangle$ (which could model an effect of
suppressing pseudo-particle contribution in the source vicinity)
in numerical calculations it becomes evident the linearly
increasing behaviour starts to form at
$\Delta/\bar\rho\sim3$---$4$. Thus, the asymptotic estimate obtained
shows the Euclidean source energy in IL gives the major contribution to
the generating functional in quasi-classical approximation if all
the coupling constant are frozen at the scale of mean instanton
size $\bar\rho$.

The linearity of static potential obtained delivers, as we believe one interesting message. 
The energy growth with the distance increasing hints
strongly at a 'great difficulty' of bringing the Euclidean colour source
into IL because the source mass (the additional contribution received
should be treated just in this way) is unboundedly increasing if the
Debye screening effects do not enter the play. It can be shown also that
corresponding integral in the color triplet (c.t.) state leads to the result
\begin{equation}
\frat{I_d}{4\pi}=4~L-l~,
\end{equation}
with color triplet mean energy
\begin{equation}
E\simeq \sigma~(4~L-l)~,~~~{\mbox (c.t.)}~,
\end{equation}
what clearly demonstrates again it is difficult to have the states with
open colour in IL. Besides, it encourages that studying the instanton
correlations and collective degrees of freedom induced by them could be
the promising way to grasp the full QCD dynamics.

If one strives to go away from the approximation of specifying the source 
orientation in the 'isotopic' space then the natural generalization of the 
results obtained takes the form ('sources of large intensity')
 $$\frat{I_d}{4\pi}=(\widetilde P_1\widetilde P_1)L+2~( \widetilde P_1 \widetilde P_2)
\left(L-\frat{l}{2}\right)+
 (\widetilde P_2 \widetilde P_2)L=
 (\widetilde P_1+ \widetilde P_2)^2~L- (\widetilde P_1 \widetilde P_2)~l~,$$
and  $e \widetilde P_1$, $e\widetilde P_2$ are considered as an
intensity of the sources with $\widetilde P=(P^1,P^2,P^3)$ denoting the 
vector of the unit normalization $|\widetilde P|=(P^\alpha P^\alpha)^{1/2}=1$ 
in colour space (the details of two body problem analysis for arbitrary 
source orientations in colour space could be found in \cite{4}$, $\cite{5}).
Surely, the result demonstrates again that it is very difficult to reveal
the states with open colour in IL and besides that there is another small 
parameter in the problem. Indeed, the deviations of vector-sources from 
the anti-parallel orientations may not be large.

Another point which is worth of a discussion here concerns the Wilson loop
behaviour. The absence of the area law for that in \cite{1} is explained
by rapid decay ($\sim \Delta^{-4}$) of the corresponding correlators in IL
and inadequate contribution of large size instantons. However, what we are
able to estimate now emphasizes an essential impact of the part of average
$G_{\mu\nu}^2$ originated by the interference term of colour source
field and stochastic (instanton) component when one determines such
observables as the interaction energy of the probe (infinitely massive)
particles. This quantity developing the linear increase at large distances
transforms into the constant gluon condensate while in the stochastic
background only. (Even in vacuum the energy required to separate two probe
colour charges from each other is finite if there are the light quarks.)
Perhaps, the physical picture could be made more transparent if other
observables are considered.

The estimate of both configurations
contribution to the generating functional (see Ref.\cite{2} for notation)
\begin{equation}
Z= \sum_{N}\frat{1}{N!}~\prod_{i=1}^{N}~
\int(d\gamma_i/\rho_{i}^5)~C_{N_c}~\widetilde\beta^{2N_c}~e^{-S}~,
\end{equation}
can be performed with the variational maximum principle. Then the
mean energy of Euclidean sources appears in the exponential factor
and may be interpreted as the suppression factor for the states
with open colour $$Z \ge e^{-E X_4 }~,$$ here we omitted the
condensate contribution together with residual Coulomb component.
Calculating the Wilson loop with this weight we find the area law
behaviour for the colour dipole in the 'isosinglet' state but 
have no transpaparent physical picture to take it as a signal of
forming the string solution.

We believe the result above shows strong interaction vacuum picture based
on the localized lumps of topological charges (which is highly successful in
describing many nonperturbative QCD phenomena and hadron phenomenology 
but at the same time was failling in treating a confinement) still have enough
credibility to become a cogent theory of QCD vacuum providing all-round
understanding of strong interaction dynamics on the same footing in lieu of a
set of individual mechanisms.
\\
\\
\\
\noindent
The authors are partly supported by STCU
\#P015c,CERN-INTAS 2000-349, NATO~2000-PST.CLG 977482 Grants.
%\newpage

\end{document}